\documentclass[aps,amsmath,amssymb,preprint]{revtex4-1}%
\usepackage{graphicx}

\begin{document}
\title{Metallic helix array as a broadband wave plate}
\author{Chao Wu}
\affiliation{Key Laboratory of Advanced Micro-structure Materials,
MOE, Department of Physics, Tongji University, Shanghai 200092,
China}
\author{Hongqiang Li}
\email{hqlee@tongji.edu.cn} \affiliation{Key Laboratory of Advanced
Micro-structure Materials, MOE, Department of Physics, Tongji
University, Shanghai 200092, China}
\author{Xing Yu}
\affiliation{Key Laboratory of Advanced Micro-structure Materials,
MOE, Department of Physics, Tongji University, Shanghai 200092,
China}
\author{Fang Li}
\affiliation{Key Laboratory of Advanced Micro-structure Materials,
MOE, Department of Physics, Tongji University, Shanghai 200092,
China}
\author{Hong Chen}
\affiliation{Key Laboratory of Advanced Micro-structure Materials,
MOE, Department of Physics, Tongji University, Shanghai 200092,
China}
\author{C.T. Chan}
\affiliation{Department of Physics, Hong Kong University of Science
and Technology, Clear Water Bay, Kowloon, Hong Kong, China}

\begin{abstract}
This study proposes that a metallic helix array can operate as a
highly-transparent broadband wave plate in propagation directions
perpendicular to the axis of helices. The functionality arises from
a special property of the helix array, namely that the eigenstates
of elliptically right-handed and left-handed polarization are
dominated by Bragg scattering and local resonance respectively, and
can be modulated separately with nearly fixed difference between
their wavevectors in a wide frequency range. The wave plate functionality is
theoretically and experimentally demonstrated by the transformation of polarized
states in a wide frequency range.
\end{abstract}

\pacs{78.67.Pt, 42.70.Qs, 42.25.Ja}

\maketitle

Manipulating the polarization of electromagnetic (EM) wave is
instrumental in both fundamental optical physics and photonics
applications. A wave plate, in the form a birefringent crystal with
specific orientation and thickness, has long been used for the
purpose of polarization control. It transforms the polarization of
electromagnetic waves by the superposition of two linearly polarized
states that are orthogonal to each other propagating inside the
crystal with different phase velocity \cite{1}. The two orthogonal
states can also be, alternatively, left-handed and right-handed
elliptically polarized (LEP and REP) eigenstates of uniaxial
bianisotropic medium \cite{2}. It is worth noting that, such a wave
plate with a certain thickness can only operate in a narrow
frequency range \cite{1,2} as the difference of phase velocity of
two polarized eigenstates is frequency dependent. A broadband wave
plate requires that, not only the difference of phase velocity but
also the axis ratios of the two polarized states must not change
throughout the whole operational band. To the best of our knowledge,
no natural or artificial medium exhibits such properties.

Metamaterials have great potential in offering a variety of novel
functionalities, such as negative refraction \cite{3}, super lens
\cite{4,5} invisibility cloak \cite{6,7}, and polarization control
via chiral route \cite{8,9,10,11,12}. The concept of metamaterial
presents a paradigm to create novel electromagnetic materials in a
form of artificial meta-atom ensemble \cite{13,14}. Crux of the
matter relies on how to tailor the electromagnetic responses and
mutual coupling between the electric field and magnetic field in
metamaterial \cite{15}. However, the most of metamaterials only
operate in a narrow frequency range due to local resonance nature.
One goal to be achieved is to find metamaterial alternatives for
related photonic devices with superior performance much beyond the
natural material limits. A helix array, as one classical
representative in chiral metamaterials, is also very unique in that
the electromagnetic properties of such kind of structure are the
collective effect between the local resonance that governs
metamaterials and Bragg scattering that governs photonic crystals
\cite{16,17,18}.

In this paper, we show that the helical symmetry of helices provides
additional degrees of freedom for tuning the dispersion branches in
different handedness, namely the REP and LEP eigenstates on the transverse
plane of metallic helix array are dominated separately by Bragg scattering
and local resonance arising from the continuous helical symmetry. The
ellipticity and difference between wavevectors of the two states can be
fixed in a wide frequency range by choosing appropriate geometric
parameters, leading to a helix solution for highly transparent broadband
wave plate. The proof-of-principle experiments in microwave regime verify
the theoretical calculations very well. The thickness-dependent character of
wave plate is also verified by the calculated and measured transmission spectra
of other model samples with different thickness. This is a first realization of a
broadband wave plate which utilizes the LEP and REP states of metallic helix array.

\begin{figure}[ptb]
\includegraphics[width=8cm]{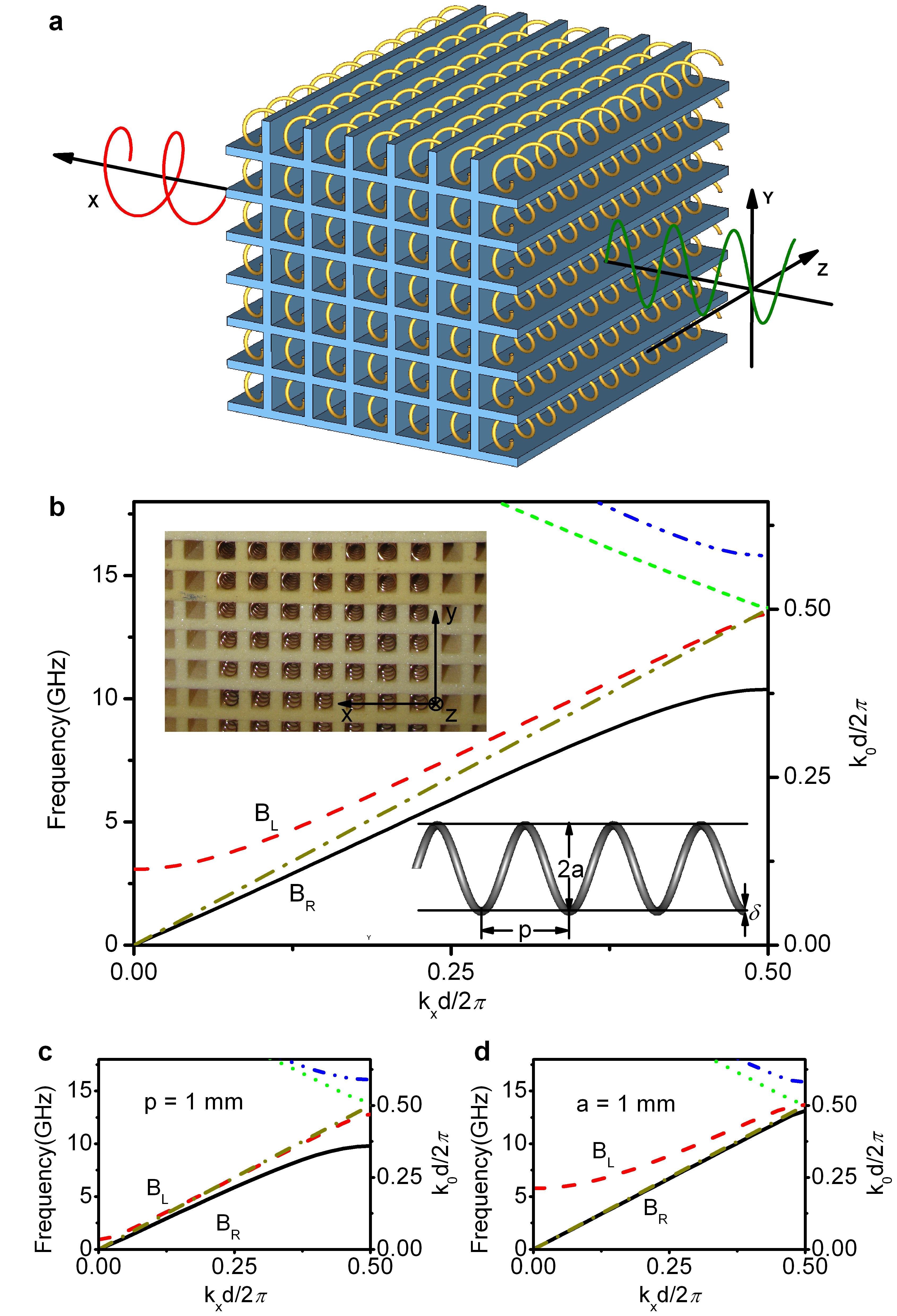} \caption{A Slab
of Helices as Wave Plate for Transversely Propagating Waves. (a) The
schematic configuration of transverse propagation through helices,
the helices are arranged in a square array; (b) the photo of the
7-layered sample and the corresponding dispersion diagram $\omega
(k_x ,k_y=k_z=0)$. The geometric parameters are the pitch $p = 4$
mm, helix radius $a = 3$ mm, wire diameter $\delta = 0.6$ mm, the
lattice constant $d=11$ mm; and the comparative results by (c)
varying pitch only to $p=1$ mm, and (d) varying helix radius only to
$a=1 $ mm, all other parameters are fixed to their respective values
in Fig. \ref{fig1}(b).} \label{fig1}
\end{figure}
\begin{figure}[th]
\includegraphics[width=8cm]{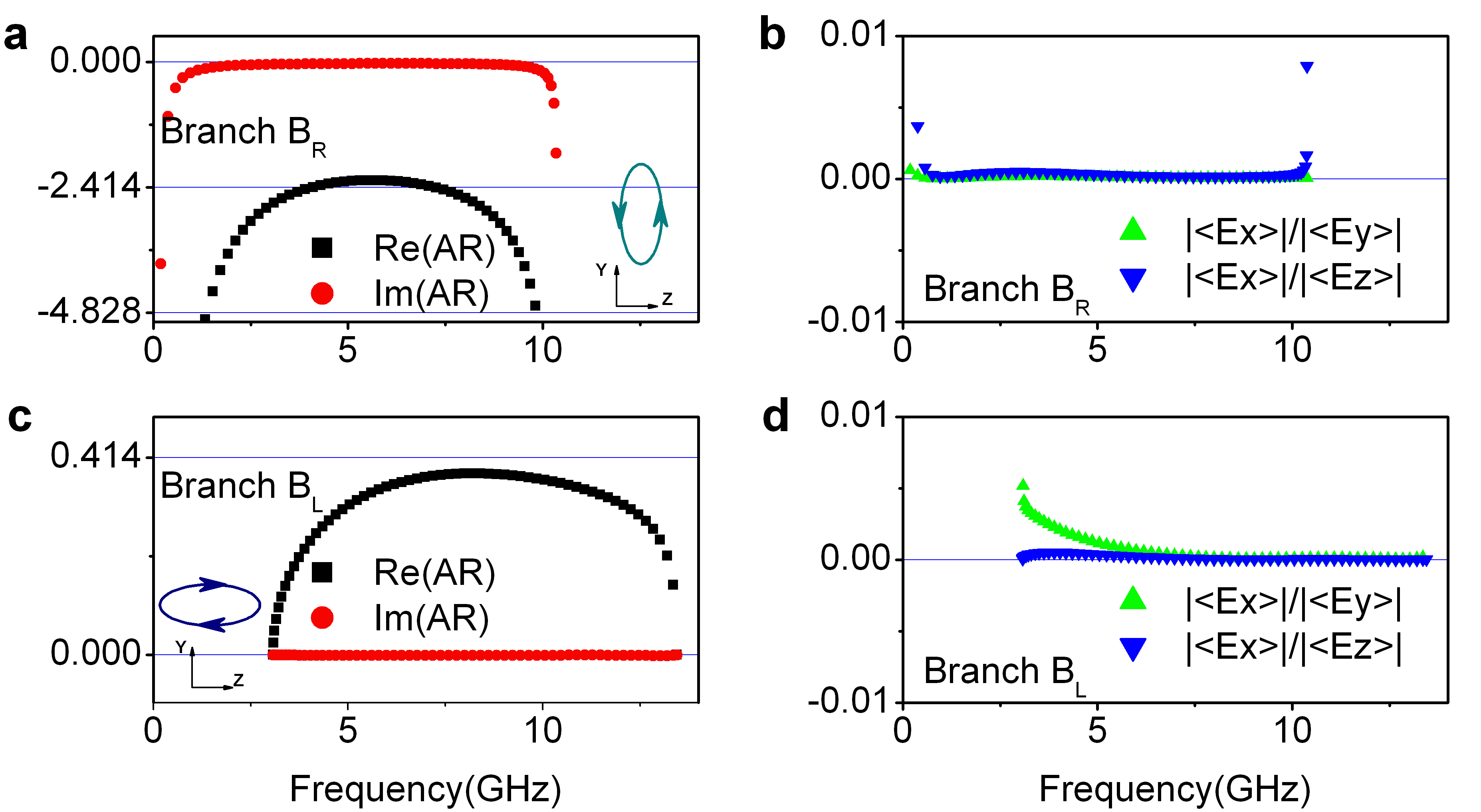}
\caption{Polarization Characters of Eigenmode. Axial ratio
$AR=\left\langle {E_y } \right\rangle /\left\langle {iE_z }
\right\rangle $, and longitudinal ratios $\left| {\left\langle {E_x
} \right\rangle } \right|/\left| {\left\langle {E_y } \right\rangle
} \right|$, $\left| {\left\langle {E_x } \right\rangle }
\right|/\left| {\left\langle {E_z } \right\rangle } \right|$ are
calculated for (a), (b) $B_R $, and (c), (d) $B_L $ modes. Black
squares for $Re(AR)$, red circles for $Im(AR)$, green
up-triangles for $\left| {\left\langle {E_x } \right\rangle }
\right|/\left| {\left\langle {E_y } \right\rangle } \right|$, and
blue down-triangles for $\left| {\left\langle {E_x } \right\rangle }
\right|/\left| {\left\langle {E_z } \right\rangle } \right|$.
$\left\langle {\ldots} \right\rangle $ denotes the
spatial average in a unit cell.} \label{fig2}
\end{figure}

Figure \ref{fig1} presents the schematic configuration on how a slab
of right-handed (RH) metallic helix array operates as a wave plate
[Fig. \ref{fig1}(a)], the photo of the 7-layered slab sample, the
corresponding dispersion diagram $ \omega (k_x ,k_y =k_z =0)$ [Fig.
\ref{fig1}(b)], and the comparative results by varying critical
parameters associated with helical symmetry and Bragg scattering
[Fig. \ref{fig1}(c) and (d)]. The geometric parameters of the sample
shown in Fig. \ref{fig1}(b) are the pitch $p = 4$ mm, the helix
radius $a = 3$ mm, the wire diameter $\delta = 0.6$ mm, and the
lattice constant $d=11$ mm. As the helical symmetry requires that a
RH helix comes back to itself after being translated by a distance
of $\Delta z$ and being rotated simultaneously by an angle of $ 2 \pi
\Delta z/p$ with $p$ being the pitch size, the field components for
an RH helix system can be expanded by functions of the form, i.e.
the helical Bloch states, as $\psi _n \left( {\rho ,\varphi ,z}
\right)=e^{ik_z z}F_n \left( \rho \right)e^{-in\varphi +2n\pi z/p}$,
where $k_z $ is the Bloch wave vector along $z$ axis, and $n$ is the
Bloch order associated with angular momentum \cite{18}. Our
calculations show that the lowest branch $B_{R}$ [black solid line
in Fig. \ref{fig1}(b)] is dictated by the degenerate $\pm 1^{st}$
orders of helical Bloch states, while the second lowest branch $B_{L
}$ [red dashed line in Fig. \ref{fig1}(b)] is dictated by the
$0^{th}$ order of helical Bloch state. The polarization characters
of eigenstates are calculated and illustrated in Fig. \ref{fig2},
shown as the longitudinal ratios $\left| {\left\langle {E_x }
\right\rangle } \right|/\left| {\left\langle {E_y } \right\rangle }
\right|$, $\left| {\left\langle {E_x } \right\rangle }
\right|/\left| {\left\langle {E_z } \right\rangle } \right|$ and the
axis ratio $AR=\left\langle {E_y } \right\rangle
/\left\langle {iE_z } \right\rangle $ of the $B_{L}$ and $B_{R}$
states, where $\left\langle {\ldots} \right\rangle $ refers to the
spatial average inside a unit cell. We see from Fig. \ref{fig2}
that, $B_{R}$ states pick up an REP character with long axis along
the $z$ axis, while the $B_{L}$ states pick up an LEP character with
long axis along the $y$ axis. These properties are essentially
decided by the helical Bloch states, since the $0^{th}$ order
helical Bloch states have their fields primarily guided along the
helix axis, while the $\pm 1^{st}$ orders of helical Bloch states
have their fields restricted on the transverse plane. The electric
field of two orthogonal elliptical states $\left| {\varphi _{LEP} }
\right\rangle $ and $\left| {\varphi _{REP} } \right\rangle $
propagating along $x$ axis can be expressed as
\begin{equation}
\label{eq1} E_{REP} =\left( {\alpha u_y +i\beta u_z }
\right)e^{i\left( {k_{REP} x-\omega t} \right)}
\end{equation}
\begin{equation}
\label{eq2} E_{LEP} =\left( {\beta u_y -i\alpha u_z }
\right)e^{i\left( {k_{LEP} x-\omega t} \right)}
\end{equation}
where $\alpha $ and $\beta $ are two different positive numbers that
determines the axis ratio of the elliptical states, $u_y$ and $u_z$
denote the unitary vectors along the $y$ and $z$ coordinate axes.

\begin{figure}[pbh]
\centerline{\includegraphics[width=8.5cm]{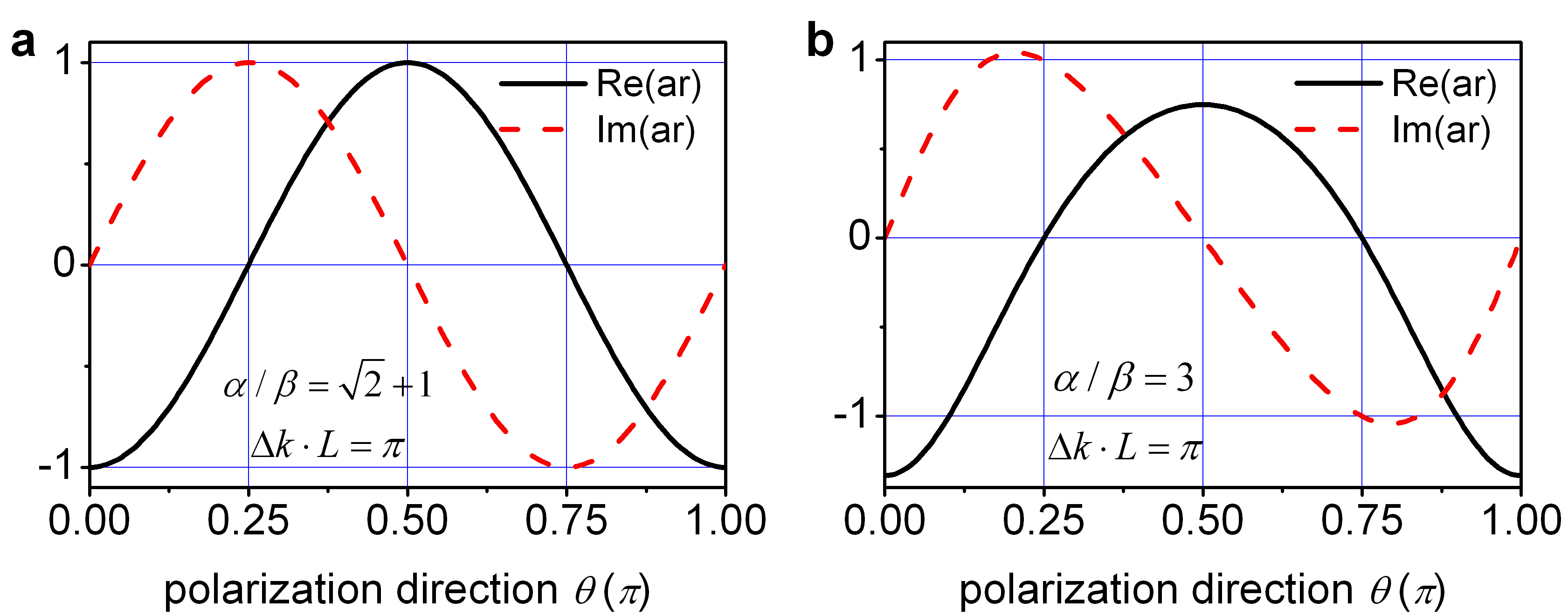}}
\caption{Polarization Transformation. Axial ratio \textit{ar} of the
transmitted waves as a function of the polarization angle of
linearly polarized incidence when the slab thickness satisfies to
$\Delta k\cdot L=\pi $and the ellipticity of the eigenstates is at
(a) $\alpha /\beta =\sqrt 2 +1$ or (b) $\alpha /\beta =3$. The ratio
\textit{ar }is defined as $ar=E_y /iE_z $, similar to the
axial ratio for eigenstates illustrated in Fig.
\ref{fig2}.}\label{fig3}
\end{figure}
\begin{figure*}[ptb]
\centerline{\includegraphics[width=15cm]{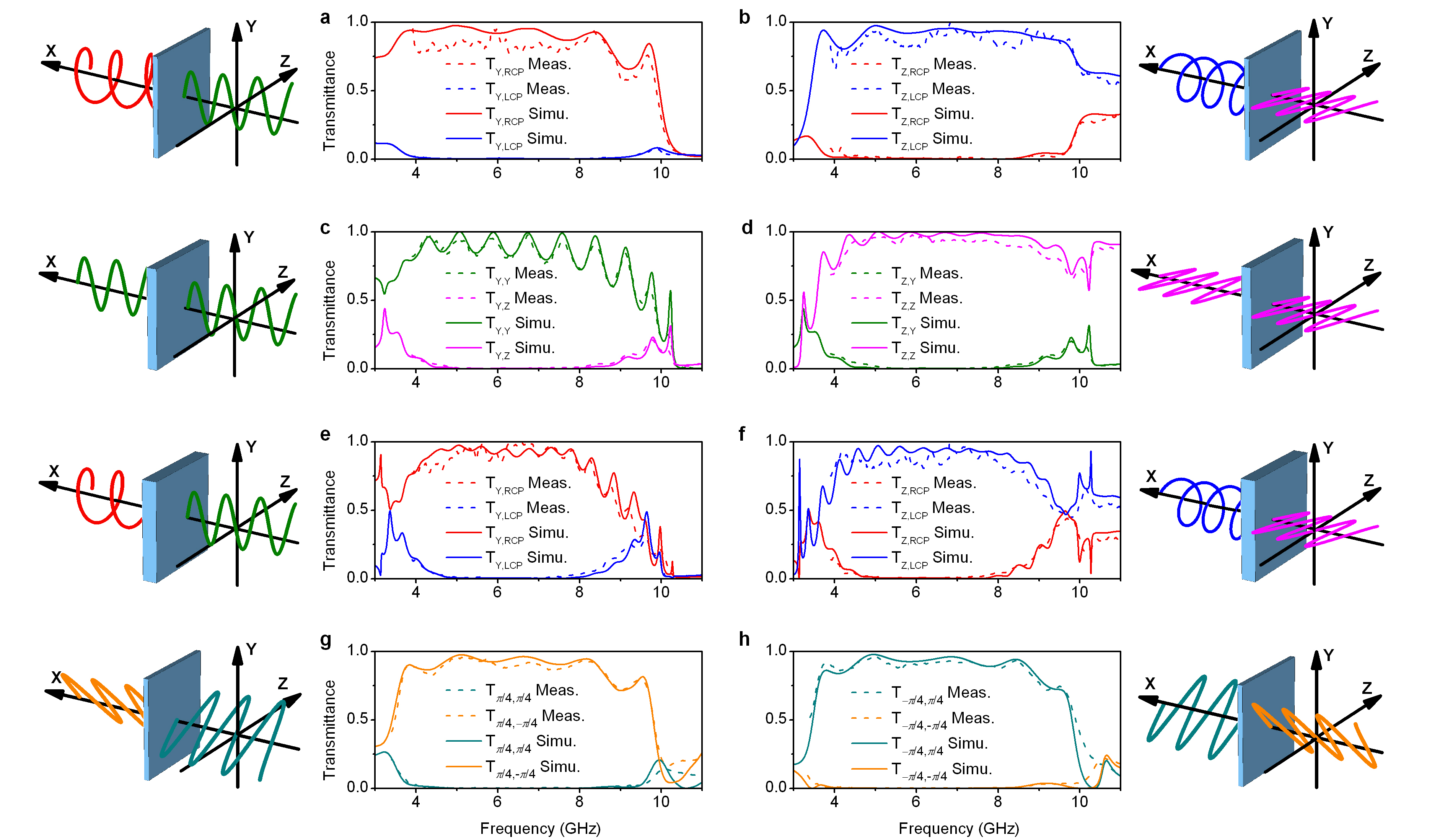}}
\caption{Transmission Spectra of Helix Samples as Wave Plate.
Transmission spectra are calculated and measured for three samples
with 7, 14 and 21 periods along the propagating direction
($x$-direction). (a) and (b), (g) and (h) for 7-period sample, (c)
and (d), (e) and (f) for 14-, 21-period samples, respectively. The
first and the second subscripts, $i$, and $j$, of the transmission
spectra $T_{i,j}$ refers to the polarized state of the incident and
the transmitted waves, respectively. The letter Y or Z denotes a
linear polarization along $y$- or $z$- direction. The symbol
\textit{$\pi $}/4 or --\textit{$\pi $} /4 denotes linear
polarization with a polarization angle at \textit{$\pi $} /4 or
--\textit{$\pi $} /4 about $y$- axis. RCP or LCP denotes a
right-handed or left-handed circular polarization. The transmission
spectra are normalized to the total power of the
incidence.}\label{fig4}
\end{figure*}

A salient feature of the band structure shown in Fig. \ref{fig1}(b)
is that the $B_{L}$ and $B_{R}$ branches are essentially linear and
parallel to each other in a wide frequency range of 3.9 GHz $\sim $
9.6 GHz. Within this band, two orthogonal eigenstates $\left|
{\varphi _{LEP} } \right\rangle $ and $\left| {\varphi _{REP} }
\right\rangle $ of the $B_{L}$ and $B_{R}$ branches pick up the same
difference $\Delta k = k_{LEP} - k_{REP} $ between their wavevectors.
This property generally holds for all directions in the transverse
plane, leading to a circular equi-frequency surface in a wide
frequency range, as if the helix array behaves as an isotropic
medium for the transversely propagating waves. We also note that the axial ratio of the in-plane
field components for the LEP/REP branch is roughly fixed within the
same frequency range as well (see Fig. \ref{fig2}). The findings are
adequate for producing a broadband wave plate which surpasses the
narrow bandwidth limitation.

The feature in dispersion diagram that two branches are in parallel
to each other in a wide frequency range is not likely to be found in
other systems. We see from Fig. \ref{fig1}(b) that the branch
$B_{R}$ is nearly linear at small values of wavevector $k_{x}$ as it
is primarily dominated by Bragg scattering. A Bragg gap is opened up
at Brillouin zone (BZ) boundary between the $B_{R}$ branch and the
fourth branch (blue dash-dot line) in the same handedness. On the
other hand, the branch $B_{L}$, in a locally resonant manner with a
cut-off frequency at $k_{x}$ = 0, becomes linear quickly with the
increase of $k_{x}$ as the $B_{L}$ states are nearly free from Bragg
scattering. If we only decrease the pitch, for example, to $p= 1 $
mm as shown in Fig. \ref{fig1}(c), the cut-off frequency of the
$B_{L}$ branch at $k_{x}= 0$  (which is primarily determined by
helical symmetry) falls down to a rather low frequency. Meanwhile
the lineshape of $B_{R }$ branch is almost not changed as it can
hardly see the chiral feature. And now the slope of the $B_{L}$
branch becomes larger than that of the $B_{R}$ branch [see Fig.
\ref{fig1}(c)]. It is also worth noting that our helix system
adopting a smaller size of radius $a$ tends to regress to an array
of thin metallic wires, and the cut-off frequency of the $B_{L}$
branch goes higher rapidly, while the $B_{R}$ branch becomes more
asymptotic to light line due to weak Bragg scattering effect as the
filling ratio of metallic helices is small. And the slope of the
$B_{L}$ branch will decrease and can be apparently smaller than that
of the $B_{R}$ branch as shown in Fig. \ref{fig1}(d).

The eigenstate analysis provides intuitive information for
interpreting the phenomena that the the $B_{L}$ and $B_{R}$ branches
can be modulated independently. A $B_{L}$ state is primarily
dictated by the $0^{th}$ order helical Bloch state with its long
axis along $z$ direction (helical axis) as shown in Fig. \ref{fig2}.
Such a state, almost free from the in-plane Bragg scattering, reveals a
resonant character arising from mutual coupling between the electric
field and magnetic field along helical axis. On the other hand, as a
$B_{R}$ eigenstate is primarily dictated by the $\pm 1^{st}$ orders of
degenerate helical Bloch states with $y$ direction as its long axis
(see Fig. \ref{fig2}), the major field components are reserved along the axial direction. And the state is heavily modulated by Bragg-scattering channels in $xy$ plane instead of mutual coupling between electric and magnetic fields along the
axial direction. As a consequence, the two branches, almost
independently modulated by the electromagnetic mutual coupling along
helical axis and the Bragg-scattering in transverse plane,
respectively, can be tuned separately \cite{19} to be essentially in
parallel to each other, given an appropriate set of geometric
parameters like those illustrated in Fig. \ref{fig1}(b). The
ellipticity of eigenstates can also be tailored much in the same way.

Consider, for example, a linearly polarized plane wave propagating
along $x$ direction through the helices with a polarization angle of
$\theta $ with respect to $y$-axis [see Fig. \ref{fig1}(a)]. The
polarization of the outgoing wave \cite{2} can be identified by its
axial ratio $ar=E_y /iE_z $ as
\begin{equation}
\label{eq3}
ar=\frac{\left( {i\alpha /\beta +\tan \theta } \right)e^{i\Delta k\cdot
L}+\left( {i\beta /\alpha -\tan \theta } \right)}{-\left( {i+\beta /\alpha
\tan \theta } \right)e^{i\Delta k\cdot L}+\left( {i-\alpha /\beta \tan
\theta } \right)}
\end{equation}
where $L$ is the slab thickness. The transmitted waves will be
transformed to be right-handed (left-handed) circularly polarized at
$\theta =0$ $(\theta =\pi /2)$ provided that the axial ratios of REP
and LEP eigenstates and the thickness of helix slab satisfy to
$\alpha /\beta =\sqrt 2 +1$ and $\Delta k\cdot L=\pi $, as shown in
Fig. \ref{fig3}(a). And the sample shown in the inset of Fig.
\ref{fig1}(b), which has 7 periods along $x$ direction, is rightly
the realization of such a wave plate with appropriately designated
geometric parameters. The wave plate utilizing REP and LEP states is
versatile in utility. We would further address that such a helix
slab can also rotate linear polarization. At the conditions $\Delta
k\cdot L=\pi $ and $\theta =\pm \pi /4$, the transmitted wave is
still linearly polarized but with rotated polarization direction
along $\mp \pi /4$. This property is independent from the axial
ratio of elliptical eigenstates as revealed in Fig. \ref{fig3} (a)
and (b). In contrast, a conventional wave plate is incapable of
implementing aforementioned two different kinds of polarization
transformation simultaneously.

According to the calculated band structure in Fig. \ref{fig1}(b),
the thinnest wave plate for linear-to-circular polarization
transformation (or vice versa) only requires 7 periods along the $x$
direction. A 7-period sample slab is fabricated by periodically
embedding the right-handed metallic helices in a polyurethane foam
slab. The polyurethane foam slab is lossless with a dielectric
constant of $\varepsilon \approx 1.01$. The sample slab contains $7
\times 60$ metallic helices (i.e. 7 periods along the $x$-direction
and 60 periods along the $y$-direction), each helix has 200 periods
along the helical axis ($z$-direction). For comparison,
finite-difference-in-time-domain algorithm is performed to calculate
the transmission spectra. Calculations show that the transmitted
waves are transformed to right-handed or left-handed circular
polarization (RCP or LCP), respectively, under $y$-polarized or
$z$-polarized incidence as shown with the solid lines in Figs.
\ref{fig4}(a) and \ref{fig4}(b). The calculated transmittance is
above 90{\%} at the most of the frequencies in the range of 3.9
$\sim $ 9.6 GHz. As the functionality of a wave plate is
thickness-dependent, we also fabricated the 14- and 21-period sample
slabs with the same geometric parameters in Fig. \ref{fig1}(b).
Following the analysis stated above, these two samples shall bring
2$\pi $ and 3$\pi $ phase difference between the LEP and REP states,
respectively. As such, the sample with 14 periods shall not change
the polarization of incident waves, and the functions of the
21-period sample shall be similar to the 7-period one. Calculated
and measured transmission spectra shown in Fig. \ref{fig4}(c) $\sim
$ (f) agree with our predictions very well. The functionality of the
7-layered sample rotating the polarization direction of a linear
polarized wave is also illustrated in Fig. \ref{fig4}(g) and (h). We
can see that the polarization angle of linear polarized wave is
perfectly transformed from $\theta =\pm \pi /4$ to $\theta =\mp \pi
/4$ as predicted.

In conclusion, we have shown that, for metallic helix system, the
dispersion branches in different handedness can be tuned
independently via separate channels. To the best of our knowledge,
this property has not been found in other systems using natural or
other artificial materials, and should be attributed to the
additional degrees of freedom provided by continuous helical
symmetry of metallic helices. By mixing the LEP and REP eigenstates,
a metallic helix array can be utilized for the realization of a
highly transparent wave plate with an ultra-wide bandwidth. The
broadband functionality is conceptually demonstrated in microwave
regime.  The helix wave plate can be versatile in utility. For
example, it can realize the perfect linear-circular polarization
transition and the polarization rotation of a linear polarized light
with an angle of 90$^\circ$ at the same broad frequency range. We also
note that it is feasible to generalize our findings to other
frequency regimes such as THz and even the infrared as well.

This work was supported by NSFC (No. 10974144, 60674778), CNKBRSF (Grant
No. 2011CB922001), Hong Kong RGC grant 600209, the National 863 Program of
China (No.2006AA03Z407), NCET (07-0621), STCSM and SHEDF (No. 06SG24).


\begin{thebibliography}{99}

\bibitem {1} M. Born, and E. Wolf, \textit{Principles of Optics:
Electromagnetic Theory of Propagation, Interference and Diffraction
of Light (7th Edition) }(Cambridge University Press, 1999).

\bibitem {2} I. V. Lindell\textit{ et al.}, \textit{Electromagnetic Waves in Chiral and BI-Isotropic Media }(Artech House, Norwood, MA, 1994).

\bibitem {3} R. A. Shelby, D. R. Smith, and S. Schultz, Science \textbf{292}, 77
(2001).

\bibitem {4} J. B. Pendry, Phys. Rev. Lett. \textbf{85}, 3966 (2000).

\bibitem {5} D. R. Smith, J. B. Pendry, and M. C. K. Wiltshire, Science \textbf{305},
788 (2004).

\bibitem {6} J. B. Pendry, D. Schurig, and D. R. Smith, Science \textbf{312}, 1780
(2006).

\bibitem {7} D. Schurig\textit{ et al.}, Science \textbf{314}, 977 (2006).

\bibitem {8} A. V. Rogacheva\textit{ et al.}, Phys. Rev. Lett. \textbf{97}, 177401 (2006).

\bibitem {9} E. Plum\textit{ et al.}, Appl. Phys. Lett. \textbf{90}, 223113 (2007).

\bibitem {10} T. Q. Li\textit{ et al.}, Appl. Phys. Lett. \textbf{92}, 131111 (2008).

\bibitem {11} N. Liu\textit{ et al.}, Nat. Mater. \textbf{7}, 31 (2008).

\bibitem {12} M. G. Silveirinha, IEEE Trans Ant. {\&} Prop. \textbf{56}, 390 (2008).

\bibitem {13} J. Pendry, Nat. Mater. \textbf{5}, 599 (2006).

\bibitem {14} N. Liu\textit{ et al.}, Nat. Photon. \textbf{3}, 157 (2009).

\bibitem {15} J. B. Pendry, Science \textbf{306}, 1353 (2004).

\bibitem {16} J. K. Gansel\textit{ et al.}, Science \textbf{325}, 1513 (2009).

\bibitem {17} J. K. Gansel\textit{ et al.}, Opt. Exp. \textbf{18}, 1059 (2010).

\bibitem {18} C. Wu\textit{ et al.}, Phys. Rev. Lett. \textbf{105}, 247401 (2010).

\bibitem {19} The $B_L$ and $B_R $ branches can be tuned seperately by varying the pitch $p$ only, or alternatively varying both the helix radius $a$ and $p$ while keep the ratio of $a/p$ to be fixed.

\end{thebibliography}
\end{document}